# Targeting Learning: Robust Statistics for Reproducible Research

Jeremy R. Coyle[1,2,**], Nima S. Hejazi[2,**], Ivana Malenica[2,**], Rachael V. Phillips[2,**], Benjamin F. Arnold[2,3,4], Andrew Mertens[2], Jade Benjamin-Chung[2], Weixin Cai[2], Sonali Dayal[2], John M. Colford Jr.[2], Alan E. Hubbard[2], and Mark J. van der Laan[2]

[**] Co-first authors: each contributed equally to this work; the ordering is alphabetical.

[1] Preva Group. 220 2nd Ave S, Seattle, WA 98104.

[2] Division of Epidemiology & Biostatistics, School of Public Health, University of California, Berkeley. 2121 Berkeley Way, Rm 5302, Berkeley, CA 94720.

[3] Francis I. Proctor Foundation, University of California, San Francisco. 95 Kirkham St, San Francisco, CA 94143.

[4] Department of Ophthalmology, University of California, San Francisco. 10 Koret Way, San Francisco, CA 94143.

## Abstract

Targeted Learning is a subfield of statistics that unifies advances in causal inference, machine learning and statistical theory to help answer scientifically impactful questions with statistical confidence. Targeted Learning is driven by complex problems in data science and has been implemented in a diversity of real-world scenarios: observational studies with missing treatments and outcomes, personalized interventions, longitudinal settings with time-varying treatment regimes, survival analysis, adaptive randomized trials, mediation analysis, and networks of connected subjects. In contrast to the (mis)application of restrictive modeling strategies that dominate the current practice of statistics, Targeted Learning establishes a principled standard for statistical estimation and inference (i.e., confidence intervals and p-values). This multiply robust approach is accompanied by a guiding roadmap and a burgeoning software ecosystem, both of which provide guidance on the construction of estimators optimized to best answer the motivating question. The roadmap of Targeted Learning emphasizes tailoring statistical procedures so as to minimize their assumptions, carefully grounding them only in the scientific knowledge available. The end result is a framework that honestly reflects the uncertainty in both the background knowledge and the available data in order to draw reliable conclusions from statistical analyses — ultimately enhancing the reproducibility and rigor of scientific findings.

# Introduction

Most writing about the crisis of reproducibility [1,2,3,4] has concentrated on issues of analysis prespecification and the changing incentives for using reproducible workflows — practices outside the imminent purview of statistics. Though the central role of statistics to reproducibility has garnered mention, there has been less focus on specific methodologies that avoid the bias inherent in traditional analytic procedures. In fact, for much of its recent history, statistical data analysis has been driven by a mindset characterized by the rote application of traditional techniques, a phenomenon now termed "cargo-cult statistics — the ritualistic miming of statistics rather than [its] conscientious practice," [2,3] which has allowed overly-specified statistical modeling choices to guide how scientific questions are answered, even when such choices result in different answers to the same research question. This practice arose naturally in the evolution of statistical data analysis: the increasing complexity of data, coupled with the widespread availability of personal computers and statistical software, made the fitting of restrictive parametric models convenient. Parametric statistical models* encode strong assumptions about the underlying processes that generated the observed data (data-generating process*; DGP), including functional forms (e.g., the outcome being linear with respect to covariates) and the error distribution (e.g., errors being normally distributed with constant variance conditional on covariates) — all potential sources of statistical model misspecification*. Unfortunately, the technically reproducible bias tied to the widespread reliance on oversimplified parametric models remains a data analytic afterthought, often left unjustified or outright ignored. Targeted Learning, an existing framework for statistical estimation and inference, provides a roadmap for learning from data — eschewing assumptions not justified by readily available background knowledge. This integrative approach takes advantage of developments in machine learning, targets the estimation toward statistical quantities that address the scientific question, and returns reliable statistical inference.

## "All models are wrong" is not a license for bias

The traditional practice, applying and interpreting parameters of low-dimensional parametric models, implicitly assumes that bias induced by misspecification is unimportant — a notion summarized in the often quoted, yet misused, maxim of G.E. Box that "all models are wrong, but some are useful" [5]. Assessing usefulness of parametric models requires substantial knowledge of the true nature of the DGP; otherwise, claims of insignificant bias rest instead on, frequently unsupported, assertions. In contrast, the practice of statistics based on realistic models (i.e., of the non-parametric variety) seeks to develop estimators* that minimize the dependence on such arbitrary assumptions. Realistic statistical models, defined as models that are large enough to contain the true DGP, should strive to only incorporate well-supported constraints on the DGP. A fundamental drive behind the epidemic of false positives plaguing scientific research, the practice of arbitrarily applying traditional statistical techniques [6,7] is made untenable when one focuses only on the knowledge truly available about the system under study. While a change in the structure and incentives of empirical research is crucial to improving the robustness of reported findings, of equal importance is the use of methodology that makes pre-specification meaningful.

## The relevant quantity to estimate is rarely a coefficient

Another problematic feature of parametric models is that model parameters often fail to reflect the scientific question of interest, even when the model assumptions are correct. The quantity of scientific interest, such as the average causal effect of a treatment on an outcome, does not depend on the entire DGP; it is represented by a specific one-dimensional summary measure of the data-generating process. Common statistical software focuses on reporting inference solely for estimates of the model parameters, such as regression coefficients in a main terms logistic regression model, with inferential procedures assuming a true (and pre-specified) model. Even if the logistic regression model happened to be correct, the average treatment effect would be represented by a function of all coefficients. The typically reported coefficient (for treatment) in a logistic regression model represents the conditional odds ratio if and only if the model is correct *and* if no interactions with treatment were included.

## Model selection leads to misleading inferences

Parametric modeling approaches, such as regression analysis, often involve model selection procedures where data-adaptive approaches are used to select a model from a sequence (e.g., stepwise regression based upon p-value thresholding). The inference is consequently made on the final model, treating it as having been pre-specified. However, unless handled in an outcome-blind setting, any interplay between the data and the final model severely biases the validity of inference. When one focuses on a coefficient, the choice of model selected also affects the interpretation of the coefficient, even if one assumes the selected model as being correct. For example, if a repeated experiment selects a different model (e.g., one including an interaction with treatment, versus one without such interactions), then the coefficient in front of the treatment variable has an entirely different interpretation. This makes estimation of the sampling distribution of an estimator and robust statistical inference difficult, if not impossible. Regression diagnostics, often relied upon heavily in such cases, may be helpful in low-dimensional settings. For example, one might plot the regression line through a scatter of points in two dimensions — however, such techniques quickly become less useful in diagnosing model misspecification with increasing dimensionality. The "guess-and-check" procedure is an underappreciated, though acknowledged, contributor to the reproducibility crisis reported in a wide range of sciences [1].

## Harnessing machine learning to address scientific questions

The Targeted Learning framework provides an approach to ensuring *methods reproducibility* [8] — "the ability to implement, as exactly as possible, the experimental and computational procedures, with the same data and tools, to obtain the same results." By allowing pre-specification of a large (and realistic) statistical model, Targeted Learning allows the admission that, given current background knowledge, very little information may be available to constrain the joint distribution of the relevant variables. Non-parametric maximum likelihood estimation*, the gold standard for estimation in parametric models, is typically impossible in high-dimensional settings due to the curse of dimensionality. Instead, data-adaptive machine learning techniques must be used to estimate the relevant features of the DGP, or the features that are required to approximate the answer to the motivating question. Moreover, when machine learning is utilized, it must focus on the the target parameter (estimand*), in order to avoid an incorrect bias-variance tradeoff. In particular, machine learning inherently optimizes the fit to the joint probability distribution of observed data, instead of the specific summary measure of scientific interest. With the advent of Targeted Learning, statistics has reached a point at which the typical interaction of data and statistical model — that same interaction that once led to potential misfitting and bias with standard regression practices — can be safely abandoned in favor of an approach that is both honest about the limitations of available knowledge and practical to implement.

Targeted Learning harnesses recent developments in causal inference, statistics and computer science to empower robust and reproducible statistical analysis. First, the field of causal inference [9] provides a rigorous framework for establishing and understanding cause-effect quantities, which enrich the interpretation of the target parameter of interest and estimate. Second, machine learning tools have thrown open the door to estimating functional relationships from data without resorting to simplistic and unrealistic models. Third, the general Targeted Learning estimation strategy integrates causal inference and machine learning to construct robust, unbiased estimators with valid inference. Targeted Learning avoids the bias-inducing interaction of the analyst and data, instead automatically guaranteeing optimal asymptotic performance and robustness in finite samples.

The Targeted Learning methodology has been utilized in a plethora of applications [10, 11, 12, 13]. A unified software environment for Targeted Learning, the *tlverse* (https://github.com/tlverse), has been built over the last few years and is designed to facilitate the principled and robust implementation of Targeted Learning. This new software ecosystem supports statistical analysis via an infrastructure that is as intuitive as classical parametric regression software. The core framework of Targeted Learning is detailed in the *Roadmap of Targeted Learning*. Both the *tlverse* and the roadmap were applied in several companion papers investigating potential causes of growth faltering in children [14–16].

# Roadmap for Targeted Learning

The roadmap for Targeted Learning provides a common framework for evaluating statistical estimation and inference problems [12, 17]. When carried out carefully, the roadmap of Targeted Learning allows the construction of efficient estimators, achieving the smallest estimation error relative to that of competing frameworks. Following the roadmap leads to a statistical estimation problem that focuses all of the information in the data towards the estimation of a specific target parameter, which represents the answer to the question of scientific interest. This represents a break from traditional approaches, which optimize the estimation globally (i.e., in terms the entire DGP). Recognizing that there is no free lunch in statistical estimation, the cost for failing to concentrate the information available in the data solely on the quantity of interest is in terms of increased statistical bias and decreased efficiency.

The Targeted Learning roadmap is generally applicable to all statistical estimation problems and consists of five steps:

1. Specify the observed data and describe the data-generating experiment;
2. Specify a statistical model representing a set of realistic assumptions about the underlying true probability distribution of the data;
3. Define a target estimand of the data distribution that "best" approximates the answer to the scientific question of interest;
4. Given statistical model and target estimand, construct an optimal plug-in estimator* of the target estimand of the observed data distribution, while respecting the model;
5. Construct a confidence interval by estimating the sampling distribution of the estimator.

Step 3 naturally builds on the field of causal inference, which defines causal models, causal quantities and methods for identification of the impact of interventions from observed data via natural experiments. Step 4 concerns the construction of *targeted maximum likelihood estimators* (TMLE), which first estimate relevant parts of the data distribution using machine learning and, subsequently, carry out a targeted updating step to tailor the machine learning fits to the target estimand. Therefore, Step 4 is split into two subparts: the machine learning step and the targeting step. In Figure 1, a simple, synthetic data example is used to illustrate the roadmap and highlight the differences between Targeted Learning and standard parametric approaches.

## Steps 1 and 2: Specify explicitly the data-generating process

Steps 1 and 2 explicitly define knowledge about the experimental design and the underlying DGP. Before analyzing the data, it is necessary to incorporate knowledge about both the experimental design and the system under study that generated the observed data. To constrain the statistical model for the data, one needs to represent all available knowledge about the DGP. Are the units independent? Were the units clustered into higher-order groups? Was the sampling unbiased or biased by design? What time-ordering generated the variables? Are there missing values, and what do we know regarding how the missingness occurred (e.g., randomly, related to measured or unmeasured covariates)? What do we know about how subjects received treatment or were exposed (e.g., randomly, related to measured or unmeasured covariates, referred to as the "treatment assignment mechanism"*)?

An understanding of how the data were generated is necessary in order to specify the relevant (groups of) observed and latent variables produced by the experiment. Collectively, this understanding of the DGP is referred to as the statistical model. Note that it contains much weaker assumptions than a parametric model about the nature of the relationship between the variables — rather, it is grounded only in what is *known* about the DGP, not what is assumed. There are several formal approaches to specifying the statistical model; the questions above provide a foundation for objectively ascertaining this information about the DGP.

In order to answer questions that can be interpreted as causal effects, one typically integrates a causal framework [18–21]. These frameworks provide tools to understand if the background knowledge, combined with the observed data, is sufficient to translate a causal question into a statistical estimand. For example, one can represent the statistical model by defining a structural causal model. A structural causal model allows us to incorporate treatment, mediators, censored and missingness variables as intervention nodes* in terms of relationships between the observed and latent variables [18]. This is often an important step for generating a target estimand that represents the answer to the scientific question of interest.

## Step 3: Define the parameter of the data-generating process that best addresses the scientific question

Many scientific questions are causal rather than associational in nature: A pharmaceutical researcher may wish to ascertain whether administering a blood pressure drug *causes* a reduction in blood pressure, or an epidemiologist may want to learn if diarrhea *causes* childhood malnutrition. The field of causal inference provides a clear framework for describing causal questions and enumerating the assumptions necessary to represent the causal quantity with a statistical estimand. A number of translational works on the topic are available [17, 22]. Importantly, causal inference motivates a range of novel statistical estimands that, under additional assumptions, admit causal interpretations. These additional causal assumptions do not affect the properties of the estimation procedure, as shown in Figure 2. We will focus directly on the statistical estimand, regardless of whether or not they are derived from a causal question, continuing the discussion of how best to estimate these from the data.

Suppose we observe, on each independently and randomly sampled unit, covariates $W$, binary treatment $A$, and an outcome $Y$. A common estimand, representing the average treatment effect (ATE), is a conditional mean difference $E_W[E(Y \mid A = 1, W) - E(Y \mid A = 0, W)]$ between treatment contrasts (the mean for $A = 1$ minus the mean for $A = 0$), for each unit's observed covariates (over all $W$). In a correctly specified linear model, this corresponds to the coefficient on $A$, and can generally be thought of as the average effect of intervention $A$ on the outcome $Y$, adjusting for covariates $W$. A minor variation of the ATE would define the target quantity as the difference in mean outcome if all units were set to intervention (i.e., $A = 1$) and the observed mean outcome, represented by the target estimand $E_W[E(Y \mid A = 1, W) - Y]$. When $Y$ is binary, this target quantity is called the population attributable risk (PAR) and it answers the question, "what would be the change in prevalence if everyone in the population received intervention $A = 1$?" [23].

A great variety of scientifically useful target quantities are defined by modifying the manner in which an intervention is assigned, possibly more flexibly than uniform assignment across all members of the target population. Consider precision medicine, which is predicated on learning how to dynamically assign interventions to different patient subgroups, in order to optimize a positive health outcome in a population. A dynamic intervention assigns treatment to individuals based on their covariates $W$, $A = d(W)$. For example, a dynamic intervention may assign older female patients a treatment that differs from older male patients, and assign all younger patients another treatment, regardless of sex. The potential outcome under this personalized intervention is represented as $Y_{d(W)}$, which we shorten to $Y_d$. The mean under the personalized intervention, $EY_d$, represents the mean outcome that would have been observed if everyone in the population had been assigned $A = d(W)$. Optimal dynamic treatments (shown in Figure 1B) are defined as the rule $d_{opt}(W)$ that optimizes (maximizes or minimizes) $EY_d$ over a user-supplied set of candidate dynamic rules. Leaving the class of treatment rules unrestricted, under the same assumptions as the ATE, the optimal rule $d_{opt}(W)$ is identified as the treatment $a'$ for which $E(Y \mid A = a', W)$ is optimized; thus, for any rule $d$, we have as mean outcome, $EY_d = E[E(Y \mid A = d(W), W)]$ [24, 25, 26]. The target estimand may be defined as the mean outcome under the optimal rule, $EY_{d_{opt}} = E[E(Y \mid A = d_{opt}(W), W)]$, possibly in contrast with the observed mean outcome $EY$ (Figure 1B). This optimal rule is estimated from the data, so one may be interested in the mean outcome under the estimated optimal rule $\hat{d}(W)$ itself (i.e., not necessarily the true optimal rule), resulting then in a data-adaptive target estimand $EY_{\hat{d}(W)}$ [27].

## Check for sufficient experimentation in observed data in order to estimate the parameter of interest without parametric assumptions

Each target estimand, such as the ATE, requires sufficient experimentation in treatment assignment — formalized as an assumption on the treatment mechanism — or *positivity*, more generally. In the case of the ATE, positivity is defined as a nonzero probability of receiving both the treatment and control conditions across all possible types of subjects in the population. That is, it must be possible to observe both levels of treatment in all covariate strata. If for some strata both treatments are not possible, the conditional mean of the outcome is undefined, since no units receiving treatment are observed within those strata. In this case, a lack of positivity means that there is insufficient information about the true value of the parameter within the given strata, thus making the ATE inestimable (i.e., not identifiable) across the entire population without extrapolation. In less extreme cases, a treatment may be exceedingly rare in certain strata, a case in which the best possible variance of the optimal estimator can be inflated.

Practically speaking, the positivity assumption for static interventions (i.e., interventions which are uniformly assigned to all subjects) is frequently violated, often when there are strong associations of a subset of the covariates and treatment, resulting in estimation instability analogous to the effect of collinearity in regression analysis. Should the set of covariates contain instrumental variables — variables that are strongly related to treatment but independent of outcome — positivity violations can prove particularly problematic. One solution is to adaptively select the estimand based on interventions truly supported by the observed data. This may be accomplished, for example, by replacing the static intervention by a rule that only sets treatment to values for which the positivity assumption is not violated. These are called "realistic rules" as they only assign treatment levels that are likely to occur for the individuals to which they are assigned, as determined by their covariates. In this manner, a target estimand that is more robustly estimable from the data may be selected over the weakly supported ATE. Regardless of the target estimand selected, Targeted Learning builds confidence intervals with widths that reflect the true degree of uncertainty. Thus, the framework provides access to a rich variety of potential target parameters, many of which will be robustly estimable from the available data and relevant for the scientific question.

The issue of positivity is rarely discussed in the context of traditional parametric modelling. This is because standard regression techniques extrapolate automatically, despite the lack of data to estimate the association of *Y* with *A* in some parts of the covariate space of *W*. Consequently, the analyst usually remains unaware of positivity issues while estimating the proposed target estimand — moreover, the width of confidence intervals fails to properly describe the true uncertainty and lack of robustness embedded in the estimation procedure. By contrast, Targeted Learning provides a means for diagnosing instances wherein positivity violations lead to problematic estimation of the estimand, without compromising statistical inference.

# Step 4: Algorithmically constructing an approximate answer to the scientific question of interest

## Optimal machine learning for an initial global fit of DGP

The target estimand for the ATE is determined by the distribution of the covariates $W$ and the outcome regression — the conditional mean $E(Y|A,W)$ of $Y$, given $(A,W)$. The probability distribution of $W$ may be estimated empirically, by simply weighting the observations equally. However, estimation of the conditional outcome regression $E(Y \mid A, W)$ in a statistical model incorporating only the (often minimal) available knowledge on its functional form is nontrivial. The machine learning literature provides many choices of flexible algorithms for fitting such a prediction function, including those from a variety of parametric models. However, the performance of any particular algorithm can vary based on the nature of the true DGP — therefore, selecting one algorithm from such a breadth of algorithms *a priori* can be challenging and problematic.

## Super Learning

The Super Learner [28] algorithm provides a template for data-adaptively selecting, from a set of machine learning algorithms, an optimal algorithm for fitting complex functions. First, we must define a measure of performance that is known to be optimized by the true function that is being estimated (e.g., the conditional mean minimizes the mean of squared residuals). Then, we must obtain a valid estimate of this performance measure across a user-specified set of candidate machine learning algorithms. This is done with cross-validation — repeatedly splitting the sample into training and testing subsets, first fitting each candidate algorithm on the training subset and then assessing performance by applying each algorithm to the test subsets, which were held-out during training. The "discrete Super Learner" or "cross-validated selector" is simply the single candidate algorithm with the best performance. The discrete Super Learner is statistically proven to perform as well as, or better than, any of its constituent algorithms [29]. Super Learner may be based on a large number of candidate learning algorithms (e.g., generalized linear models, lasso [30], neural networks, regression trees [31], random forest [32], BART [33]), including variations of the same algorithm with different choices of tuning parameters. Theory and extensive simulations suggest that Super Learners constructed from large sets of candidate machine learning algorithms offer performance gains relative to Super Learners based on smaller sets of candidate algorithms; thus, Super Learner accommodates and benefits from a large diversity of adaptive and smooth learning techniques [12,29]. Beyond the discrete Super Learner, the ensemble Super Learner variant is constructed by combining ("ensembling") several candidate algorithms into one optimal algorithm by selecting the best weighted combination of the candidates.

While Super Learner may be used to optimally estimate the outcome function $E(Y|A,W)$, the target estimand is typically a specific summary measure of $E(Y|A,W)$, averaged over $\{A,W\}$, such as those defined in Step 3 of the roadmap. In such cases, plugging in the Super Learner estimate of $E(Y|A,W)$ into the target estimand equation is insufficient. In particular, simply plugging the Super Learner fit into the target estimand equation generates a substitution estimator* that is overly biased relative to its variance, and fails to converge to a normal sampling distribution, which is required to obtain a measure of uncertainty for the estimate. Despite doing a better job in terms of approximating the true outcome function relative to using a parametric model, Super Learner alone does not provide valid statistical inference, including confidence intervals and p-values. In addition, methods based on resampling, such as the bootstrap [34], also fail to produce valid inference when applied to the Super Learner. The *targeting step* described below resolves the incompatibility in using data-adaptive estimation and obtaining valid inference by updating the outcome function estimate returned by the Super Learning such that is targeted towards the estimand of interest. This update reduces the bias that remains in the Super Learning fit with respect to the target estimand, and guarantees asymptotic normality which is needed for statistical inference. Indeed, one of the great conveniences of this approach is that resultant estimators allow a simple method for deriving robust statistical inference, one that is applicable to a vast array of possible target estimands of scientific interest.

## Targeted Maximum Likelihood Estimation

There is a vast literature on targeted maximum likelihood estimation [12,13,35]; we encourage the interested reader to consult this literature for complete and detailed mathematical justifications of the methodology. The targeted maximum likelihood approach builds on a rich history of estimation of causal effects in semi-parametric models, the development of which owes a great deal to James Robins and collaborators [36–38]. Most of the original methodological developments concentrated on the approach of estimating equations [39,40]. Here, we concentrate instead on the core elements of the more recently developed framework of Targeted Learning, aiming to develop intuition for how the approach provides a powerful roadmap for deriving estimators with optimal asymptotic and robust finite-sample performance. The most common approach to deriving such estimators relies on asymptotic normality and semiparametric efficiency theory [41]. The resultant estimators are maximally efficient (i.e., lowest possible variance) within the class of regular estimators* and allow for the construction of asymptotically valid confidence intervals and hypothesis tests.

An essential concept in statistical estimation theory is *asymptotic linearity*: the ability to approximate the difference of an estimator and the true parameter value as an average of independent and identically distributed random variables. If an estimator can be represented in an asymptotically linear form, then a straightforward approach to both estimating the sampling variance and sampling distribution is accessible. More technically, the difference between an asymptotically linear estimator and the target estimand can be written (asymptotically) as an average of independent and identically distributed random variables: $\hat{\Psi} - \Psi \approx \frac{1}{n} \sum_{i=1}^{n} IC(O_i)$, where $IC(O_i)$ is called the *influence curve* (IC). As described above, the Super Learner plug-in estimator does not accommodate this expression, making valid statistical inference impossible to attain. For any given target estimand, there exists a set of possible estimators, each with a corresponding IC. Among these, one IC, termed the *efficient influence curve* (EIC), has the lowest sampling variance. Any estimator corresponding to this EIC is said to be *efficient*, meaning that it has the lowest possible variance.

With the EIC in hand, two general techniques for constructing efficient asymptotically linear estimators of the target estimand exist: the estimating equation estimator (EEE) (as well as its one-step approximation, OSE), and targeted maximum likelihood estimation (TMLE). TMLE can be understood as an update to an initial estimator (e.g., Super Learner) of the data distribution. The form of the update is specific to the target parameter, based on using maximum likelihood estimation in a clever, low-dimensional (e.g., one-dimensional) parametric model, with the initial estimator acting as an offset. Intuitively, the update step tailors the initial estimator fit to the target estimand. The tuning parameter of the low-dimensional parametric model is such that a small change implies a maximal change in the target estimand, so that the maximum likelihood fit of this tuning parameter essentially fits the target estimand. This corresponds with the tuning parameter of the parametric model having score equal to the EIC. By maximizing the likelihood, and thereby solving the EIC score equation*, the TMLE update step reduces bias and minimizes sampling variability, thus utilizing the observed data in the best manner possible for learning the target estimand.

Though both EEE and TMLE are based upon the EIC, several important theoretical reasons exist for preferring TMLE [12]. Firstly, as a plug-in estimator, the TMLE will always respect bounds on the estimate implied by the statistical model and target estimand (e.g., always within [0, 1] when the target estimand is a probability), making the TMLE more robust than the EEE. Secondly, by extending the clever parametric model in the updating step, one can force the TMLE to respect other equations beyond the EIC equation. Based on this principle, TMLEs with additional properties, including enhanced finite-sample performance, doubly robust inference [42] [43], or guaranteed improvement over a specified estimator [12] [13], have been constructed.

## Step 5: Report uncertainty

Since the TMLE is asymptotically linear with the EIC as its influence curve, its variance can be estimated with the sample variance ($\sigma_n^2$) of the estimated EIC values $EIC_n(O_i), i = 1, \ldots, n$, scaled by sample size $n$. To improve the variance estimator, we use the sample variance over a validation sample with EIC fitted on the corresponding training sample, averaged across sample splits [44]. An asymptotically valid 95%-confidence interval may be constructed as $\psi_n^* \pm 1.96 \frac{\sigma_n}{\sqrt{n}}$, analogous to a confidence interval based on the sample mean, with $\psi_n^*$ representing the constructed TMLE-based estimate.

Figure 1A "4. Construct Estimator" section compares the fits of the original Super Learner and the small perturbation induced by the TMLE update. The result is a slightly wider separation of the lines corresponding to the estimates $E(Y \mid A = 1, W)$ and $E(Y \mid A = 0, W)$, implying that residual confounding from the Super Learner-based estimate causes an underestimate of the treatment impact. For this estimand, the principal concern is the probability that the confidence intervals contain the true value (coverage), shown in Figure 1A "5. Form Inference" section. The linear regression-based substitution estimator with the coefficient estimate on treatment and its standard error is clearly biased — its confidence interval failing to include the true ATE. The Super Learner-based estimator provides an improvement in terms of bias, but is limited since no formal inference is possible (thus, just the estimate is shown). Finally, the TMLE-based estimate both reduces bias and allows construction of confidence intervals using the estimated EIC. When one does not know the truth, the interval around the regression-based estimate appears preferable, due to its smaller width of the confidence interval suggesting enhanced power. This highlights a central problem with traditional regression approaches: in order to properly account for bias, confidence intervals must be significantly wider than those built by standard regression-based inferential techniques.

## Demonstration of approach

In order to compare the approaches, we need to evaluate their behavior in a setting where the true value of the statistical estimand is known, and in which we can generate repeated samples. Consequently, we conducted a Monte Carlo simulation by sampling repeatedly from the same DGP and estimating the target parameter on each sample. In this case, we simulated from the same DGP used in Figure 1, where we considered a sample size of 100 subjects and the treatment was not helpful for all subjects. From this simulation, we generated the sampling distribution of our estimation strategies (Figure 3A). The generalized linear model (GLM) is heavily biased in this synthetic, but realistic example, with a skewed sampling distribution, whereas TMLE has only minimal bias and is close to being normally distributed — an important property for obtaining valid inference. We can also compare these estimators with a number of performance metrics (Figure 3B). We see that, compared to GLM, TMLE has lower bias, variance, and mean squared error. In addition, it has close to nominal coverage (i.e. 95% confidence intervals cover the true parameter value 95% of the time), despite the fact that its confidence intervals are only trivally wider than those of the GLM.

## Targeted Learning in real-world data science

The first paper on Targeted Learning was published in 2006 [35], with Super Learning [28] a year later. Many of the early publications relevant to this topic focused on developing estimators for new target estimands and on making the estimators more robust [12] [45]. There is a growing literature on both the performance of Targeted Learning relative to competing estimators and its application across myriad disciplines. For instance, several recent publications demonstrated the superior performance of Super Learner in prediction relative to other candidate approaches using cross-validation and test datasets [46,47,48]. Additionally, several recent works illustrated the surpassing performance of Targeted Learning methods in terms of lower estimation error and improved inferential accuracy, using data-generating mechanisms inspired by randomized controlled trials [49,50], observational studies [51], and "challenging" simulations popularized in the causal inference literature [51]. Other studies compared the results of implementing competing methods with access to information external to the available data, yet still observing superior performance from Targeted Learning [10,11,52,53]. Since 2016, Targeted Learning-based results have been published in numerous manuscripts assessing observational studies, many concerning scenarios in which standard techniques fail — for example, in cases of estimating the impact of longitudinal interventions [54,55] [56,57] [58][59, 60]. Further, on account of its ability to increase estimation efficiency, TMLE has been used to analyze data from randomized controlled trials [49, 50], the gold standard for causal inference.

The Targeted Learning roadmap provides a template for the construction of efficient plug-in estimators of target estimands for data distributions for any type of structure and statistical model. To highlight the flexibility of the framework, we survey several types of scientifically impactful but complex target parameters.

## Multiple time point interventions

In clinical settings, treatment decisions are often dynamic and incorporate real-time information about observed units. Consider an observational study evaluating available strategies aimed at minimizing microvascular complications by controlling glucose level across time. In this setting, individualized treatment rules that intensify the treatment for controlling glucose when a unit's glucose level crosses a cutoff might be of interest, with each cutoff representing a particular dynamic treatment strategy. Since the intervention is not defined by a singular drug administration event, estimands defined by single time point interventions cannot accommodate the scientific question of interest. Under a causal framework, one can define counterfactuals and the post-intervention distribution as the distribution of the data in which all interventions had been carried out on all subjects at all time points $t = \{1, ..., k\}$. Of course, this scenario is ideal and cannot be observed in practice, since one cannot roll back time and modify observed intervention patterns. Nonetheless, the target causal quantity is defined accordingly as the probability of not having a microvascular complication at time $k+1$ under the post-intervention distribution. TMLE-based estimators have both been developed for target estimands based on complex longitudinal data structures and applied to real-world datasets via their R package implementations [61, 62, 63, 64, 65, 66].

## Stochastic treatment regimes

The target estimands presented thus far assume the treatment of interest takes on one of only a small set of values (i.e., binary or categorical) and involve static or dynamic treatment rules as functions of only baseline covariates. In many interesting scenarios, the assigned intervention is made a deterministic function $d(A, W)$ of covariates $W$ and the natural (possibly continuous) treatment $A$ (e.g., being assigned to engage in more exercise than one already performs). For instance, one might want to estimate the outcome of an experiment in which we modify the current exposure by a relative amount of the observed value — for example, reduction of exposure by 10%. Such interventions are referred to as stochastic, or "feasible interventions", and correspond to randomly drawing, for a given $W$, the treatment by first examining the observed treatment $A$ and then evaluating the deterministic rule $d(A, W)$. Importantly, the deterministic rule $d(A, W)$ can be designed to account for covariates $W$ in such a way that positivity violations can be avoided entirely.

Consider a study in which one seeks to ascertain the effect of reducing surgical operating time for cancerous tumor extraction by 20 minutes on mortality or cancer relapse [67]. As the natural operating time for an individual is needed to define the intervention, the goal becomes estimation of the counterfactual* mean outcome under a *stochastic* rule $d(A, W)$ = A-20, or the average rate of clinical deterioration if all patients' operating times were reduced by 20 minutes.

TMLEs have been developed for target estimands of mean outcomes under stochastic interventions, utilizing machine learning and TMLE updating of estimators of the treatment mechanism and outcome regression [68, 69]. Stochastic interventions, and the TMLEs of features of the corresponding post-intervention distributions, naturally generalize to the above multiple time point setting.

## Network-dependent data

Suppose one observes a community of units over time, collecting treatment status, outcome and covariates. For each unit, suppose one knows which other units in the community potentially influence it (e.g., friends and family of the subject). The DGP and likelihood at a given time point depend on the observed past of that unit and its connections. By replacing the observed treatment distributions across units and time by a desired community-level intervention, a post-intervention distribution of the data can be constructed. A target quantity could be defined as the expected final outcome across units under this intervention. In light of the COVID-19 pandemic, consider a U.S. county in which a social distancing order was implemented. The counterfactual disease risk for a given individual could only be assessed based both on whether a given individual obeys the county-level order and on the participation of the unit's immediate network (e.g., family, friends, neighbors). TMLE-based procedures have been tailored for causal queries and corresponding target estimands concerning a community of interconnected individuals with a known network structure [70, 71, 72].

## Cluster-level interventions and hierarchical data

Often, individual-level outcomes are correlated via cluster-level exposure or social, biological or geographical factors. In such settings, one might seek to estimate the impact of an exposure randomly, or naturally, assigned at a cluster level. Consider a hierarchical DGP where a community is randomly selected from a target population, with sampling of units taking place within each cluster (e.g., individuals from a specific hospital or school). The causal impact of a cluster-level intervention may be formally defined and identified from observed data with a non-parametric causal model, and a TMLE-based estimate of the resulting target estimand could be constructed under the assumption that the communities are independent. Such a formulation allows for many sources of dependence between individuals within a cluster, including direct transmission of the outcome and influence of the covariates of a given individual on the outcome of another [73].

## Adaptive sequential randomized trials

Consider a randomized trial in which new patients are enrolled over time; randomized to treatment arms, possibly conditional on the baseline covariates and followed until the outcome of interest is observed. For example, one might focus on adaptive surveillance methods for the COVID-19 pandemic which aim to allocate tests in a manner that identifies new infections sooner, and monitors changes in epidemic status with greater accuracy and precision. At any point in time when the treatment (e.g., COVID-19 PCR test) must be assigned to a patient, (i) data on previously tested patients, (ii) the patient's network of person-to-person contacts and (iii) the patient's medical history and comorbidities could be used to learn and adjust the randomization probabilities of receiving the test such that the patients facing highest risk for infection are tested. Continuously updated randomization probability estimates may be used to randomize initial rounds of treatment to new, incoming patients, as well as to randomize subsequent treatment decisions to already enrolled patients. Alternatively, for long time-series data, one might instead be interested in learning, and consequently adapting, treatment decisions based on the evolution of a single time-series [13]. At any point in time in an ongoing adaptive sequential trial, one may wish to estimate the mean outcome under the current best estimate of the optimal dynamic treatment rule instead of the more arduous true optimal dynamic treatment rule. TMLEs have been developed for sequential adaptive randomized trials, using machine learning to learn the optimal dynamic treatment and cross-validation to estimate the mean outcome under the estimated optimal rule [74].

# Robust methods for reproducible data science

Statistics provides a template to precisely translate a real-world application into a statistical estimation problem in terms of a formulation of the data, the statistical model of the data distribution, and the target estimand. Honest formulation of the statistical model requires incorporating concrete knowledge and making only limited assumptions on the data distribution, such as conditional independence assumptions and bounds, if known. Causal and censored data models allow one to first define the question of interest in terms of a "full-data" estimand*, so that causal identification results may be leveraged to generate a corresponding statistical target estimand of the observed data distribution under specified, but generally non-testable, assumptions.

To translate real-world data with all of its complexities into actionable information, one needs *a priori*-specified estimation procedures with valid statistical inference — as can be constructed through the roadmap of Targeted Learning. This roadmap provides a template for the construction of targeted, plug-in machine learning algorithms that are asymptotically optimal and flexible enough to optimize finite-sample behavior, through specification of the Super Learner algorithm library and subsequent TMLE updating step. Targeted Learning accommodates state-of-the-art advances from a variety of disciplines — causal inference, machine learning and statistical theory — into a unified approach for statistical learning. By leveraging computationally demanding machine learning algorithms, Targeted Learning can, in turn, benefit from the continued increases in available computing power.

## The Future of Targeted Learning

The Targeted Learning framework has been informed by a combination of rigorous theory and empirical testing. Statistical theory is used to establish that an estimator is asymptotically linear and normally distributed, is optimally efficient, and has confidence intervals with the desired asymptotic coverage of the true parameter value. For any given statistical formulation of the estimation problem, the performance of an estimator and its inference can be evaluated across simulated datasets, ideally in a setting where the analyst lacks access to the true estimand values. In this way, a laboratory can be manufactured in which state-of-the-art estimators may be iteratively developed and objectively evaluated with both theoretical and practical benchmarks. In our experience, such an approach has proven instrumental in continuously improving the finite-sample performance of the Targeted Learning framework, relying upon new theoretical developments that can be naturally incorporated in the flexible TMLE framework.

To develop these methods to their fullest potential, we work with simulated and real data in collaboration with biologists, medical researchers, government agencies, epidemiologists, and other companies. We also have made a centralized effort to construct a collection of Targeted Learning software packages that work well together and foster future software development in a unified fashion. A free and open source handbook (https://tlverse.org/tlverse-handbook/) on the use of the *tlverse* software ecosystem is in preparation [75].

# Glossary

| Term | Definition |
| --- | --- |
| Data-generating process (DGP) | The true mechanism that generated the observed data, with the corresponding data-generating probability distribution which produces the observed samples that were collected. |
| Statistical model | A set (family) of probability distributions that could describe the data-generating process. Note that the true data-generating process is unknown. |
| Full-data and full-data model | The data one would have observed in the ideal (impossible) experiment, and the set of possible probability distributions of the full-data random variable. In a causal model, the full data includes the counterfactual values of the outcome (i.e., potential outcomes) under all treatment/exposure conditions. |
| Counterfactual | A contrary-to-fact value said to arise from hypothetically imposing an intervention on a system represented by a structural causal model. For example, the potential outcome $Y_a$ is a counterfactual that arises from a hypothetical intervention that sets the treatment $A$ to level $a$. |
| Model misspecification | A scenario in which the statistical model, which is postulated to contain the distribution describing the data-generating process, fails to actually contain the corresponding true data-generating distribution. |
| Parametric statistical | A family of probability distributions indexed by a finite set of |

| model | parameters. For example, a linear model traditionally assumes the outcome is a linear function of covariates plus a normally distributed error term with constant variance. Its parameters are the coefficients on the covariates and the variance of the error term. |
|---|---|
| Non-parametric or infinite-dimensional statistical model | A family of probability distributions that cannot be indexed by a finite set of parameters. That is, the set of parameters indexing this family of distributions is infinite-dimensional. Most often, when making minimal assumptions, the data-generating process cannot be defined by a finite set of parameters, making the set of parameters infinite-dimensional. For example, if all we know about the data-generating process is that we have access to $n$ independent and identically distributed (i.i.d.) samples, then the statistical model for the data-generating process is a non-parametric statistical model. |
| Treatment assignment mechanism | The mechanism by which treatment assignment decisions are made, which can be defined as a conditional probability distribution of treatment (e.g., $P(A|W)$ in our ATE example). Formally, this is a component of the data-generating process. In a classical randomized controlled trial, $P(A|W) = 0.5$. |
| Censoring mechanism | The mechanism by which units are censored in the observed data, which can be defined as a conditional distribution of censoring variables, given full data. For example, $P(\Delta|A,W)$ captures how outcomes $Y$ became censored as a function of their baseline $W$ and treatment $A$, where $\Delta$ is a binary indicator equalling $0$ for subjects with missing outcomes and $1$ otherwise. |
| Estimator | A function of the sample of observations (that is, a function of the random variables) that generates estimates. |
| Estimate | The realized value of an estimator, or a function of the realized observations. |
| Target estimand or target parameter | A function of the true (unknown) data-generating process that one is interested in estimating, and represents the mathematical formulation of the motivating question of interest. |
| Maximum likelihood estimation | The most common method for estimating parameters in a finite-dimensional model (i.e., parametric statistical model). As the name implies, such estimates are generated by finding a set of parameter estimates that maximize the likelihood function of the observed data. |
| Score equation | The gradient (i.e., multi-variable generalization of the derivative) of the |

| | log-likelihood function of the data with respect to the parameter(s). This equation provides information on the degree of change resulting from very small perturbations of the parameter values. |
|---|---|
| Regular estimator | A class of estimators that converge in distribution to some limit distribution even if one samples from a slightly perturbed data distribution. Such estimators, if also asymptotically linear, accommodate inference by way of their asymptotic convergence to a Normal distribution. |
| Plug-in (substitution) estimator | An estimator that generates an estimate of the true parameter value by "plugging in" estimates of relevant parts of the data-generating distribution into the parameter mapping. This method is commonly referred to as the plug-in principle. For example, "plugging in" targeted Super Learner fit of the conditional mean under $A = 1$ and $A = 0$ generates an estimate of the average treatment effect. |

# Acknowledgments

This research was financially supported by a global development grant (OPP1165144) from the Bill & Melinda Gates Foundation to the University of California, Berkeley, CA, USA.

# Roadmap of Targeted Learning

## A. ROADMAP FOR THE AVERAGE TREATMENT EFFECT (ATE)
### Comparison of Standard Approach to Targeted Maximum Likelihood Estimation

**1. DESCRIBE DATA**

$n = 100$ subjects

For each subject, pre-treatment covariates ($W$), treatment ($A$), and outcome ($Y$) vectors were measured

$W$ pre-treatment covariate

$A$ indicating whether subject received treatment ($A = 1$) or not ($A = 0$)

$Y$ continuous post-treatment outcome

Subjects were sampled independently from each other and from the same population distribution $P_0$

$O_1, \ldots, O_n \overset{iid}{\approx} P_0$

**2. SPECIFY STATISTICAL MODEL**

*Standard Approach*

Parametric statistical model

Does not contain $P_0$, the DGP (i.e., misspecified model)

True data-generating process (DGP)

*Targeted Maximum Likelihood Estimation*

Realistic semiparametric or nonparametric statistical model

Defined to ensure $P_0$ is contained in model

**3. DEFINE ESTIMAND**

Additional assumptions are required to interpret this estimand as causal

$\Psi$ is a function that takes as input $P_0$ and outputs the answer to the question of interest

*What is the average effect of treatment on outcome?*

$$\Psi(P_0) = E_0\big(E_0[Y|A=1,W] - E_0[Y|A=0,W]\big)$$

The **assumption of positivity** is required to estimate of this quantity from the data. That is, it must be possible to observe both levels of treatment for all strata of $W$.

**4. CONSTRUCT ESTIMATOR**

*Standard Approach*

*Generalized Linear Model (GLM)* to estimate

$\mathbf{Y} = \beta_0 + \beta_1 \mathbf{A} + \beta_2 \mathbf{W} + \epsilon$

Estimated coefficients are biased

Cannot detect heterogeneity in treatment effect

Treatment ($A$): 0, 1

Estimator: GLM, Super learner (step 1), TMLE (steps 1 and 2)

*Targeted Maximum Likelihood Estimation*

*TMLE* implements a two-step procedure

1. initial estimation of $E_0[Y|A,W]$ with super (machine) learning
2. targeting towards optimal bias-variance trade-off for $\Psi(P_0)$

TMLE estimates are unbiased and doubly robust

*To have legitimate estimate and inference, statistical model must be (1) correctly specified (i.e., contain $P_0$) and (2) selected a priori, before looking at the data.*

**5. FORM INFERENCE**

*Standard Approach*

Inference (such as $p$-value and confidence interval) is misleading and erroneous

True ATE, $\Psi(P_0)$

Point Estimates and 95% Confidence Intervals

*Targeted Maximum Likelihood Estimation*

Targeting (step 2) improves estimate and makes inference possible

Trustworthy inference obtained with efficient influence function

## B. ROADMAP FOR THE OPTIMAL TREATMENT EFFECT (steps identical to ATE omitted)

**3. DEFINE ESTIMAND**

*What is the effect of optimal individualized treatment on outcome?*

$$\Psi(P_0) = E_0\big(E_0[Y|A = d_{\text{opt}}(W), W]\big) - E_0[Y]$$

$d_{\text{opt}}(W)$ is a decision rule that tailors treatment assignment to the subject (based on their characteristics $W$) to maximize their expected outcome.

$E_0[Y]$ is the mean outcome under the observed treatment assignment.

The optimal intervention assigns treatment $A = 1$ if $W \geq 5$ and $A = 0$ if $W < 5$

Treatment ($A$): 0, 1, Optimal

**4. CONSTRUCT ESTIMATOR**

Step 1 of TMLE requires estimation of the conditional additive treatment effect (CATE)

$E_0(Y|A=1,W) - E_0(Y|A=0,W)$,

to learn the decision boundary for optimal rule $d_{\text{opt}}(W)$ to assign treatment $A$

Decision boundary for $d_{\text{opt}}(W)$

Estimated, True

Treatment is beneficial / Treatment is harmful

**5. FORM INFERENCE**

True effect of optimal individualized treatment, $\Psi(P_0)$

Super learner, TMLE

Point Estimates and 95% Confidence Intervals

**Figure 1.** Roadmap of Targeted Learning using a synthetic data example that applies generally to health- or science- related studies where subjects receive a treatment or exposure. (A) shows the roadmap for the average treatment effect (ATE), comparing the standard approach, a Generalized Linear Model (GLM), to Targeted Maximum Likelihood Estimation (TMLE). The statistical interpretation of the ATE is the average of the difference in means between treated and control groups, averaged across covariate strata. (B) shows the roadmap for the optimal individualized treatment effect, removing the steps that are identical to (A). The interpretation of the optimal individualized treatment effect is the average of the difference between the mean conditional outcome given the optimal individualized treatment and covariate $W$, and the outcome under the observed treatment assignment. (B) omits the standard approach from consideration, since a GLM cannot be used for estimation in this setting.

The graphs in (A) "2. Specify Statistical Model" and "3. Define Estimand" sections include curves of the true mean outcome under the observed treatment assignment and the optimal treatment assignment, respectively, given the baseline covariate. Also, both the points in both graphs constituted the observed data. In both (A) and (B), the graphs included in the "4. Select Estimator" sections display the estimated lines of best fit according to the corresponding estimation strategy.

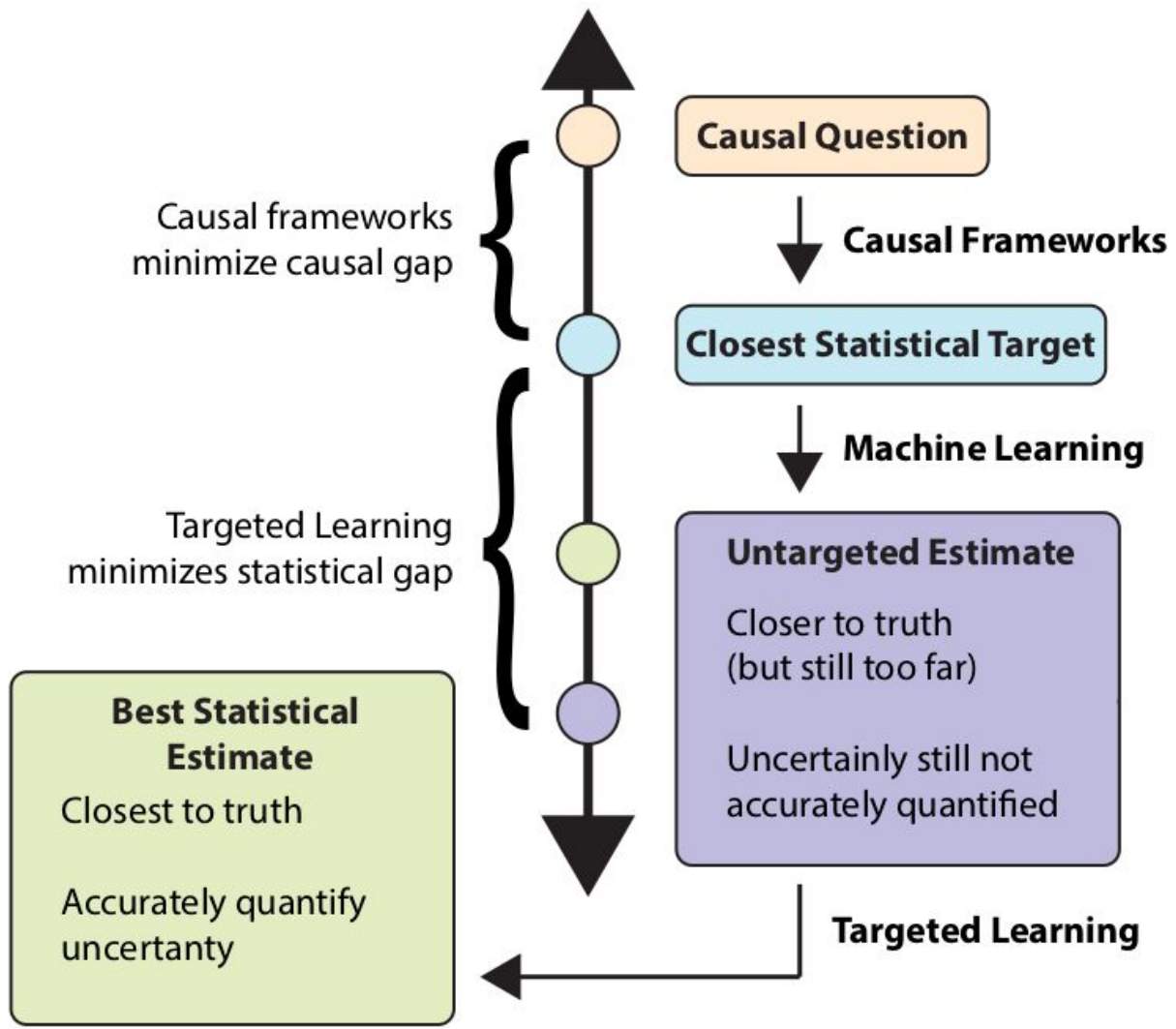

**Figure 2**. Differentiating the role of causal inference from the role of statistical estimation. The field of causal inference provides a powerful and expressive framework for describing causal questions and enumerating the assumptions necessary to represent the causal quantity with a statistical estimand, which is defined in terms of the observed data. The observed data may not meet all assumptions required in order interpret the estimate causally (e.g. no unmeasured confounding). The causal gap represents the difference between the observed data and the so-called "full-data" — the information needed in order to ascertain causality. The assumptions required to interpret the estimate as a causal relationship do not change the statistical model or the statistical estimation problem. Thus, *estimator properties are not affected by the causal gap*. The statistical gap concerns Targeted Learning estimators, and this gap takes into consideration the statistical properties of an estimator. Here, the statistical properties include bias (closeness to the truth) and inference (quantification of uncertainty), but others should also be taken into consideration (e.g., efficiency).

# Simulation Results

**A.** Sampling Distributions of Estimators

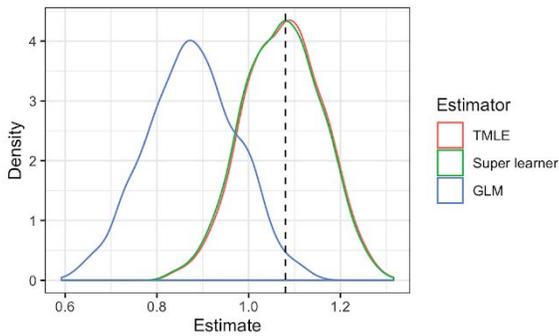

**B.** Performance Metrics of Estimators

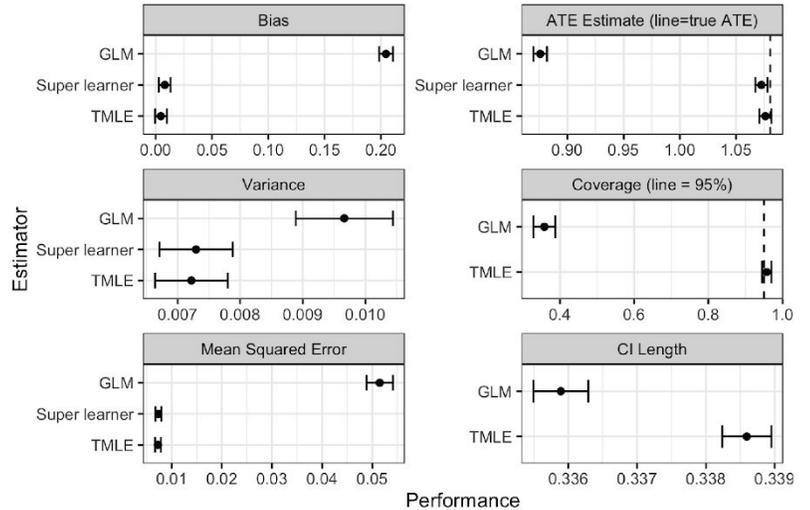

**Figure 3**. Results from 1,000 simulation iterations sampled from the same synthetic data-generating process presented in Figure 1. (A) shows the sampling distributions (the distribution of the estimated values across the simulation iterations) of the three estimators (TMLE, Super learner, and GLM). The dashed line indicates the true estimand value. (B) shows the statistical performance of the estimators. The dashed lines indicate the true estimand value.